\documentstyle[12pt,prc,aps,psfig]{revtex}
\tighten
\begin{document}
\draft

\title{Microscopic calculations in asymmetric\\
nuclear matter}
\author{D. Alonso and F. Sammarruca}
\address{Physics Department, University of Idaho, Moscow, ID 83844, U.S.A}
\date{\today}
\maketitle
\begin{abstract}
A microscopic calculation of the equation of state for asymmetric nuclear 
matter is presented. We employ realistic nucleon-nucleon forces and operate within
the Dirac-Brueckner-Hartree-Fock approach to nuclear matter. The focal point of this paper is a  
(momentum-space) G-matrix which properly accounts for the               
asymmetry between protons and neutrons. 
This will merge naturally into the development of an effective interaction
suitable for 
applications to asymmetric nuclei, which
will be the object of extensive study in the future.
\end{abstract}
\narrowtext

\section{Introduction}

 Nuclear matter is an idealized  uniform infinite system of protons and neutrons
under their mutual strong forces and without electromagnetic interactions.
 Symmetric nuclear matter (that is, equal densities of protons and
neutrons) has been studied extensively. The so-called conventional
 approach to nuclear matter goes back to earlier works by Brueckner and others [1-6] and is known 
 as the BHF (Brueckner-Hartree-Fock) theory. 
 During the 1980's, the Dirac-Brueckner-Hartree-Fock (DBHF) approach
was developed [7-9]. The 
 break-through came with the observation that the DBHF theory, unlike the 
 conventional one,
 could describe successfully the saturation properties of nuclear matter, that
 is, saturation energy and density of the equation of state (EOS).
The DBHF method adopts realistic nucleon-nucleon (NN) interactions and contains features 
of the relativistic theory. It characterizes the nuclear mean field by 
strong, competing scalar and vector fields that together account for the binding
of nucleons as well as the large spin-orbit splitting seen in nuclear states.
The DBHF framework is a reliable, as well as feasible, 
microscopic method to describe effective interactions in the nuclear medium.

Concerning asymmetric nuclear matter,                                         
systematic empirical investigations to determine
its saturation properties have so far not been done.
From the theoretical side, some older studies can be found in Refs.\cite{BC68,Siem}.
Interactions adjusted to fit properties of finite nuclei, such as those based
on the non-relativistic Skyrme Hartree-Fock theory \cite{B+75} or the relativistic mean field
theory \cite{ST94}, have been used to extract 
 phenomenological EOS.                       
Generally, considerable model dependence is observed among predictions based    
on different EOS \cite{O+98}.
Variational calculations of asymmetric matter have also been reported \cite{LP81}.
In Ref.~\cite{FLW}, a Lorentz invariant functional of the baryon field operators is defined
to project Dirac-Brueckner nuclear matter results onto the meson-nucleon vertices of an effective
density-dependent field theory. This is then applied to asymmetric matter and finite nuclei 
in Hartree calculations \cite{HKL}. 

 We use realistic NN 
forces and operate within the DBHF framework. Similar calculations have been done     
in Ref.~\cite{Oslo}.                                                                   
The purpose of our paper, however, is to produce a 
self-contained report, including all relevant formulas and nuclear matter results,
and, most important, to examine the effect of the asymmetry on the in-medium NN interaction.
This work will 
serve as a baseline for the future applications we plan to pursue. 
These include reaction studies using effective interactions based on asymmetric
matter suitable for scattering on finite nuclei, and 
calculations of neutron radii and neutron skins. Concerning properties of astrophysical 
implications, we calculate the pressure in symmetric matter and neutron matter at high densities
(up to about five times saturation density), and compare with recent experimental constraints
obtained from analyses of nuclear collisions \cite{nscl}, with the extracted pressures being the highest 
recorded under laboratory conditions.

Recently, much interest has developed around the study of 
highly asymmetric nuclei (in particular, extremely neutron-rich nuclei, and 
halo nuclei). If approved for construction, the Rare Isotope Accelerator (RIA) will map the 
limits of nuclear existence and allow the study of the unique nuclear systems which
populate those boundaries.
Thus, it is important and timely to develop microscopic effective interactions
which can account for the asymmetry between proton and neutron densities. 

In the self-consistent approach, a calculation of nuclear matter properties 
 yields, at the same time, a convenient parametrization of the density dependence
in the form of nucleon effective masses. This information can then facilitate the 
calculation of the scattering matrix at positive energies, and finally of an effective interaction
suitable, for example, for proton scattering on asymmetric nuclei.                      
This interaction will be ``isospin dependent", in the sense of being different for
the $nn$, $pp$, or $np$ cases.                                                      

 Here, we will first describe our DBHF calculation of asymmetric matter (Section II). In Section III we will present and discuss 
 results for the EOS at various levels of asymmetry, which confirm the findings from Ref.~\cite{Oslo}. We then compare high-density pressure calculations to most recent experimental observations. In the context of non-relativistic 
calculations, we also consider a popular potential based on chiral perturbation theory.
 In Section IV, we explore the 
 sensitivity of the G-matrix at various positive energies to increasing values of the neutron excess parameter.
Conclusions and plans for future applications are presented in Section V.

\section{Description of the calculation}

Asymmetric nuclear matter can be characterized by the neutron density, 
$\rho_n$, and the proton density, $\rho_p$. It is also convenient to define the total density
$\rho = \rho_n + \rho_p$ and the asymmetry (or neutron excess) parameter
$\alpha = \frac{ \rho_n - \rho_p}{\rho}$. 
Clearly, $\alpha$=0 corresponds to symmetric matter, and 
$\alpha$=1 to neutron matter.                       

In terms of $\alpha$ and the average Fermi momentum, related to the total density in the usual way, 
\begin{equation}
  \rho =\frac{2 k_F^3}{3 \pi ^2} , 
\end{equation}
the neutron and proton Fermi momenta can be expressed as 
\begin{equation}
 k^{n}_{F} = k_F{(1 + \alpha)}^{1/3}      
\end{equation}
and 
\begin{equation}
 k^{p}_{F} = k_F{(1 - \alpha)}^{1/3} , 
\end{equation}
 respectively.
 
We use the Thompson relativistic three-dimensional reduction 
of the Bethe-Salpeter equation. The Thompson equation is applied to nuclear matter in
strict analogy to free-space scattering and reads, in the nuclear matter rest frame,                 
\begin{eqnarray}
&& g_{ij}(\vec q',\vec q,\vec P,(\epsilon ^*_{ij})_0) = v_{ij}^*(\vec q',\vec q) \nonumber \\[4pt]
&& + \int \frac{d^3K}{(2\pi)^3}v^*_{ij}(\vec q',\vec K)\frac{m^*_i m^*_j}{E^*_i E^*_j}
\frac{Q_{ij}(\vec K,\vec P)}{(\epsilon ^*_{ij})_0 -\epsilon ^*_{ij}(\vec P,\vec K)} 
g_{ij}(\vec K,\vec q,\vec P,(\epsilon^*_{ij})_0)  \nonumber \\[4pt]                     
\end{eqnarray}                    
where $ij$=$nn$, $pp$, or $np$, and the                                      
asterix signifies that medium effects are applied to those quantities.
In Eq.~(4),                                  
$\vec q$, $\vec q'$, and $\vec K$ are the initial, final, and intermediate
relative momenta, and $E^*_i = \sqrt{(m^*_i)^2 + K^2}$. 
The momenta of the two interacting particles in the nuclear matter rest frame have been expressed in terms of their
relative momentum and the center-of-mass momentum, $\vec P$, through
\begin{eqnarray}
\vec P = \frac{\vec k_{1} + \vec k_{2}}{2} \\
\vec K = \frac{\vec k_{1} - \vec k_{2}}{2}
\end{eqnarray}                    
The energy of the two-particle system is 
\begin{equation} 
\epsilon ^*_{ij}(\vec P, \vec K) = 
e^*_{i}(\vec P, \vec K)+  
e^*_{j}(\vec P, \vec K)   
\end{equation} 
 and $(\epsilon ^*_{ij})_0$ is the starting energy.
 The single-particle energy $e_i^*$ includes kinetic energy and potential 
 energy, see below.            
The Pauli operator, $Q$, prevents scattering to occupied states.            
 To eliminate the angular
dependence from the kernel of Eq.(4), it is customary to replace the exact
Pauli operator with its angle-average. 
Detailed expressions for the Pauli operator in the case of
two different Fermi momenta are given in Appendix {\bf A}. It is also customary to 
introduce an average center-of-mass momentum \cite{bh5}. Definitions and details
are also provided in Appendix {\bf A}. 

With the definitions
\begin{equation} 
G_{ij} = \frac{m^*_i}{E_i^*(\vec{q'})}g_{ij}
 \frac{m^*_j}{E_j^*(\vec{q})}             
\end{equation} 
and 
\begin{equation} 
V_{ij}^* = \frac{m^*_i}{E_i^*(\vec{q'})}v_{ij}^*
 \frac{m^*_j}{E_j^*(\vec{q})}             
\end{equation} 
 one can rewrite Eq.(4) as
\begin{eqnarray}
&& G_{ij}(\vec q',\vec q,\vec P,(\epsilon ^*_{ij})_0) = V_{ij}^*(\vec q',\vec q) \nonumber \\[4pt]
&& + \int \frac{d^3K}{(2\pi)^3}V^*_{ij}(\vec q',\vec K)
\frac{Q_{ij}(\vec K,\vec P)}{(\epsilon ^*_{ij})_0 -\epsilon ^*_{ij}(\vec P,\vec K)} 
G_{ij}(\vec K,\vec q,\vec P,(\epsilon^*_{ij})_0) \nonumber \\[4pt]                       
\end{eqnarray}                    
which is formally identical to the non-relativistic equation. We make use of
the definitions Eqs.(8-9) throughout this paper, which is why all formulas involving
the opertor $G$ appear identical to their non-relativistic equivalent.

The goal is to determine self-consistently the nuclear matter single-particle potential   
which, in our case, will be different for neutrons and protons. 
To facilitate the description of the numerical procedure, we will use a schematic
notation for the neutron/proton potential (while the corresponding detailed expressions
are reported in Appendix {\bf B}). 
We write, for neutrons,
\begin{equation}
U_n = U_{np} + U_{nn}  
\end{equation}
and for protons
\begin{equation}
U_p = U_{pn} + U_{pp}  
\end{equation}
where each of the four pieces on the right-hand-side of Eqs.(11-12) depends on the appropriate 
G-matrix ($nn$, $pp$, or $np$) from Eq.(10).                                           
Clearly, the two equations above are coupled through 
the $np$ component and so they must be solved simultaneously. Furthermore, 
the G-matrix equation and Eqs.(11-12)  
are coupled through the single-particle energy (which includes the single-particle
potential). So we have three coupled equations to be solved self-consistently.
As done in the symmetric case \cite{db2}, we 
parametrize the single particle potential for protons and neutrons (Eqs.(11-12)) in terms of
two constants, $U_{S,i}$ and $U_{V,i}$, (the scalar and vector potential) through       
\begin{equation}
U_i(k_i) = \frac{m^*_i}{E^*_i}U_{S,i} + U_{V,i}.
\end{equation}
For the purpose of facilitating 
the connection to the usual non-relativistic framework \cite{bh5}, it is customary to 
redefine $U_{S,i}$ and $U_{V,i}$ in terms of                                      
two other constants defined as 
\begin{equation}
m^*_i = m_i + U_{S,i} 
\end{equation}
and 
\begin{equation}
U_{0,i} = U_{S,i} + U_{V,i}. 
\end{equation}
The subscript ``$i$'' signifies that these parameters are different for protons and
neutrons. 
Starting from some initial values of $m^*_i$ and $U_{0,i}$, the G-matrix equation is 
 solved and a first approximation for $U_{i}(k_i)$ is then obtained. This solution is 
again parametrized in terms of a new set of constants, 
and the procedure is repeated until convergence is reached.     
The effective masses for neutrons and protons, $m^*_i$, obtained through this procedure
are shown in Figs.~(1) as a function of the Fermi momentum at various levels
of asymmetry.

Finally, the energy per neutron or proton in nuclear matter is calculated from 
\begin{equation}
\bar{e}_{i} = <T_{i}> + <U_{i}>. 
\end{equation}
(See Appendix {\bf C} for details.)
 The EOS, or energy per nucleon as a function of density, is then written as
\begin{equation}
    \bar{e}(\rho_n,\rho_p) = \frac{\rho_n \bar{e}_n + \rho_p \bar{e}_p}{\rho} 
\end{equation}
or 
\begin{equation}
    \bar{e}(k_F,\alpha) = \frac{(1 + \alpha) \bar{e}_n + (1-\alpha) \bar{e}_p}{2}. 
\end{equation}

\section{Properties of the Equation of State} 

 The NN potential used in this work is the  relativistic OBEP from Ref.\cite{Mac89} 
which uses the Thompson equation and the pseudo-vector coupling for the $\pi$ and $\eta$
mesons. 

 The EOS as obtained from our DBHF calculation is displayed in Fig.~2 (upper panel), as a function of 
$k_{F}$ and for values of $\alpha$ between 0 and 1. The symmetric matter EOS saturates 
at $k_{F}\approx 1.4 fm^{-1}$ with a value of 16.7 MeV, in good agreement with the 
empirical values.   
For the compression modulus of saturated symmetric matter, defined as 
\begin{equation}
 \kappa = k_F^2 \frac{\partial ^2 \bar{e}(k_F)}{\partial ^2 k_F}|_{k_F=k_{F}^{(0)}}     
\end{equation}
we find a value 
of 233 MeV. This is in excellent agreement with                                   
the recent empirical determination of 225$\pm$15 MeV \cite{PD02}.

 As the neutron density increases (the total density remaining constant),                 
 the EOS becomes increasingly repulsive and the minimum shifts 
towards lower densities. 
As the system moves towards neutron 
matter, the ``energy well" gets more and more shallow, until, for $\alpha$ larger 
than about 0.8, the system is no longer bound.             

 We also show for comparison the EOS based on the BHF calculation, see Fig.~2, lower panel.                      
Also in this case we have performed a self-consistent calculation based on the same realistic 
force. However, no medium modifications are included in the potential to account for the 
proper Dirac structure of the nucleons in nuclear matter. As a consequence of that, saturation 
of symmetric matter is obtained at a much higher density, a well-known problem with the 
conventional approach. 
 
 In Fig.~3, we take a different look at the EOS. There, we plot the quantity 
   $\bar{e}(k_F,\alpha) - \bar{e}(k_F, 0)$ versus $\alpha^2$. Clearly, the behaviour is linear, that is
\begin{equation}
   \bar{e}(k_F,\alpha) - \bar{e}(k_F, 0) = e_s \alpha ^2,                                      
\end{equation} 
or parabolic versus $\alpha$.                                                                   
This linear behaviour,                                                                           
shared with the non-relativistic predictions, see lower panel of 
Fig.~3, is reminiscent of the asymmetry term in the familiar semi-empirical mass
formula. 
By reading the 
 slope of each line, we can then predict the nuclear symmetry energy, defined as          
\begin{equation}
e_s = \frac{1}{2} \frac{\partial ^2 \bar{e}(k_F,\alpha)}{\partial ^2 \alpha}|_{\alpha=0}. 
\end{equation} 
 This is shown in Fig.~4, where the solid curve is the prediction 
from the DBHF model and the dashed corresponds to the BHF calculation. The DBHF prediction 
at saturation density is about 30 MeV.                                                              
The behaviour of the symmetry energy is most controversial at high densities, where even   
  the trend depends strongly on the type of interaction used \cite{BL}.
 Energetic reactions induced by heavy neutron-rich nuclei have been 
proposed as a mean to obtain crucial information on the high-density behaviour of the 
nuclear symmetry energy and thus the EOS of dense neutron-rich matter \cite{BL}. 
Analyses of collision dynamics have been done at the National Superconducting Cyclotron
Laboratory                 
 to extract EOS-sensitive observables such
as the elliptic flow \cite{nscl}. In this way, empirical constraint has been obtained for the pressure of symmetric and neutron matter. 

The pressure of neutron-rich matter is defined in terms of the energy/particle as 
\begin{equation}
P(\rho, \alpha)=\rho^2 \frac{\partial \bar{e}(\rho,\alpha)}{\partial \rho}               
\end{equation} 
(We notice that Eq.~(20) facilitates the 
calculation of the pressure for any $\alpha$.)

In Fig.~5, we show the pressure obtained from the DBHF-based EOS for symmetric 
matter and neutron matter.           
The empirical constraints have been provided by the authors of Ref.~\cite{nscl}. 
The area within the dashed lines indicates EOS consistent with the elliptic flow measurements 
reported in that work. 
For the case of neutron matter, two different parametrizations for the symmetry energy are assumed in the 
analysis. The two pressure contours in the bottom panel of Fig.~5 correspond to the weakest (lower contour) and the strongest (higher contour) density dependence 
for $e_s(\rho)$ proposed by Prakash {\it et al.} \cite{Prak}.          

It is fair to say that DBHF predictions produce a reasonable amount of repulsion. 
In contrast, relativistic mean field theories (RMFT) tend to generate 
too much pressure \cite{nscl}, as can be expected from the higher compression modulus usually predicted by RMFT. 
While very repulsive EOS such as obtained from RMFT as well as weakly repulsive EOS (compression modulus of 167 MeV)
are ruled out by the analysis, 
EOS which become softer at densities higher than three times saturation density (possibly due to a phase transition to quark matter) are not excluded.

\subsection*{ Non-relativistic calculations with a high-precision chiral potential}

The predictions shown above are typical of meson-exchange potentials with 
moderate strength in the tensor force (as indicated by a deuteron $D$-state probability
of about 5\%). 
In-medium predictions, however, can be very  sensitive to the nature of the chosen 
two-body force, and can be so even if the interactions under consideration predict very similar  
 free-space observables. 

 Recently, a new generation of NN potentials based on 
chiral perturbation theory ($\chi$PT) has begun to emerge
\cite{EM,Epel}. These interactions may become                       
  increasingly popular in the future as a 
 way to describe low-energy processes in terms of effective 
degrees of freedom while respecting the symmetries of QCD.
Thus we felt it would be interesting to consider one of these potentials 
as well in calculations of symmetric and asymmetric matter.                      
 In this subsection, we show basic EOS properties for the chiral potential 
of Ref.~\cite{EM}, which is quantitative in its description of NN data up to
300 MeV. 

Chiral potentials are non-relativistic in nature, and thus unsuitable
for a DBHF calculation. For comparison, we include predictions from a modern
high-precision meson-exchange potential (the CD-Bonn from Ref.~\cite{CD}).
 (The symmetry energy for various high-precision potentials has been calculated in Ref.~\cite{Oslo2}.)
CD-Bonn, which uses pseudoscalar coupling for the pion and has been developed within the 
framework of the non-relativistic Lippman-Schwinger equation, will also be 
used, as it is appropriate, in a conventional Brueckner calculation. 

When evaluationg these predictions, one must keep in mind that consideration of many-body
forces is a crucial issue in non-relativistic calculations (whereas some three-body forces are 
effectively included in the DBHF framework \cite{Bro87}).                 
The chiral potential, in particular, requires the inclusion of 
three-body forces for consistency within its order. However, we are interested in a baseline comparison 
between the two interactions                                                                  
as it will reveal differences, if any, originating at the 
two-body level. 
In fact, the free-space predictions of the two potentials under our consideration are of very similar quality. However, the momentum
structure of chiral potentials can be quite different than the one of conventional 
meson-exchange potentials, with the chiral interactions being ``softer'' due to cutoffs applied 
to eliminate high-momentum components, as required by the nature of the chiral expansion.
Such differences are likely to show up in G-matrix calculations.                                       

EOS properties from the chiral potential in comparison with CD-Bonn are displayed in 
Figs.~6-8.                                                                                                 
Generally, the behaviour is similar to the one of a low tensor force potential, such as Bonn A, when used  
in a non-relativistic context (the chiral potential predicts a deuteron $D$-state probability of
4.93\%). 
 Although similar at moderate densities, the two sets of predictions start differing considerably 
for higher values of the Fermi momentum. The best overview is provided by the symmetry energy, Fig.~8,
where we see the two curves being very close to each other up to about saturation density, after which the 
chiral curve seems to get a bit steeper, but than crosses over to become softer at the higher densities.

Due to the cutoff applied to the chiral potential which suppresses   
momentum components higher than 460 Mev/c \cite{EM}, it's legitimate to ask up to which value of the 
Fermi momentum its predictions can be deemed reliable. As a guideline for this estimate, we will take
the case of symmetric matter, which is obtained from the formulas in the appendices setting 
the proton and neutron Fermi momentum equal to each other. Evaluation of 
the energy/particle requires integration over the relative momentum (denoted as $K_0$ in Appendices {\bf B}-{\bf C}) 
up to $k_F$, which is therefore the maximum value of $K_0$.                  
With the  cutoff of 460 Mev/c being equivalent to 2.3 fm$^{-1}$, 
the chiral potential predictions are realistic for the values of $k_F$ shown                   
in Figs.~6-8.

\section{In-medium interactions } 

The G-matrix calculated from Eq.~(10) has the dependence (for a given energy and
momentum),
\begin{equation}
 G^{\beta, \beta '}_{ij} =                      
 G^{\beta, \beta '}_{ij}(k_F, \alpha) , 
\end{equation} 
where $\beta$ is an appropriate set of quantum numbers, and $ij$ refers to the 
type of nucleons.                                                                    
We now wish to solve Eq.~(10) for scattering states. Envisioning a physical situation 
where protons are the 
incident particles, we then have $pp$ and $pn$ possibilities, which are calculated 
separately with the appropriate effective masses and Pauli blocking operator. 

As is well known since the birth of nuclear physics, the $pp$ and $pn$ interactions
in the same state 
would be identical except for charge dependence. However, in the 
asymmetric medium, these interactions are physically different due to the different Fermi momenta
of protons and neutrons. In Fig.~9, the neutron and proton Fermi momenta from Eqs.~(2-3) are shown
versus the asymmetry parameter $\alpha$ for fixed (average) Fermi momentum. 
                
To which extent the asymmetry in the proton/neutron ratio in a nucleus will affect
proton-nucleus scattering is equivalent to the question: how sensitive is the 
interaction to the asymmetry degree of freedom? 
We can certainly get insight into this issue from the (infinite nuclear matter) 
G-matrix, which,                                                           
upon proper mapping into coordinate space and application of the local density approximation, will yield the 
effective interaction for finite nuclei \cite{S1,S2}. When ``folded'' over the nucleus, the interaction
yields the optical potential  used to describe elastic scattering, or, in the case of inelastic scattering,
it becomes the transition potential connecting the initial state to the target excited states.

At present, scattering calculations on halo nuclei are being done with 
phenomenological interactions where the isospin asymmetry is accounted for through
different parametrizations for $pp$ and $pn$ \cite{Gupta}. The Melbourne group \cite{Amos} 
has produced a very comprehensive set of (non-relativistic) proton-nucleus scattering calculations, including some exotic 
nuclei. However, no attempt is made to account for the asymmetry in the interaction.
Clearly, additional calculations would be helpful. 

For this sensitivity study, we have selected one S-wave and one P-wave with T=1,
so that $pp$ and $pn$ cases are both existing and can be compared. The $^{1}S_{0}$ wave
will give us insight into the central force, while the 
 $^{3}P_{1}$ wave contains strong sensitivity to the spin-orbit component.
 (The central and spin-orbit pieces of the effective interaction play the most important role in  
  proton-nucleus elastic scattering and inelastic scattering to natural-parity transitions.) 

In Figs.~10-11,  we show the real and imaginary parts of the (on-shell) $^{1}S_{0}$  matrix element
at three different energies. In each case, the quantity is plotted versus the 
(average) Fermi momentum, for three values of $\alpha$. 
 The same legend applies to Fig.~12-13 for 
 the $^{3}P_{1}$  matrix element. 

In general, the (smaller) imaginary part is much more sensitive to the degree of asymmetry.
The presence of an imaginary part is determined by the occurrence of a singularity in 
the scattering equation, Eq.~(10). That occurrence will depend on the kinematics and the 
value of the Fermi momenta. We see for instance that at 50 MeV 
the imaginary part vanishes at higher densities for larger values of $\alpha$. This is particularly 
striking in the $pp$ case, due to the drastic lowering of the proton 
Fermi momentum as seen from Fig.~9.

It is interesting to notice how the $\alpha$-dependence shows up differently in the $pp$/$pn$            
cases.               
The dependence on $\alpha$ comes in through the $p$/$n$ effective masses (see Fig.~1) and
of course Pauli blocking, which acts differently on $pp$ or $pn$ states (see Appendix {\bf A} for details).
Pauli blocking in $pp$ 
 scattering is controlled by just the proton Fermi momentum, whereas 
$pn$ scattering is controlled by both 
the neutron and proton Fermi momenta, which vary as seen from Fig.~9.                              
As a result, the scattering probability 
in the $pn$ states exhibits overall a weaker dependence on $\alpha$.                       

We would expect a similar trend for in-medium $pp$ or $pn$ cross sections, a very useful concept for       
transport model calculations \cite{Bali}, where both the mean field and nucleon-nucleon collisions
play an important role. 
For instance, we would expect 
in-medium $pp$ cross sections to be more sensitive than $pn$ cross sections to increasing asymmetry.
By the same token, $nn$ cross sections would be strongly Pauli-blocked in neutron-rich matter. 
 We will look into this aspect in a future work.

Finally, as the energy increases, the details of Pauli blocking become much less important, and so does
 the $\alpha$-dependence, although                                                                              
  the overall density dependence remains strong. 
(At higher energies, relativistic density-dependent effects, which are included, become the most important.) 
 Experiments at low energy would be best to reveal sensitivity of the interaction to
the asymmetry between neutron and proton distributions.

\section{Summary and conclusions } 

We have presented a microscopic calculation of the equation of state of
nuclear matter when protons and neutrons have different Fermi momenta.
The calculation is self-consistent and parameter-free, in the sense that no parameter
of the NN force is adjusted in the medium. 

As expected, the single-particle energy moves up to less attractive
values to merge with the neutron matter equation of state when the proton density
approaches zero.
The dependence of the EOS on the neutron excess parameter is clearly linear 
 as a function of $\alpha ^2$. We make predictions for the nuclear symmetry 
energy and observe a large discrepancy between the relativistic and the non-relativistic 
predictions at high density. Our fundings are in agreement with those from
Ref.~\cite{Oslo}. 

A relatively simple way to relate the microscopic EOS directly to 
structural properties of finite nuclei is the use of a mass formula 
\cite{O+98}, where 
the ``volume" term is directly related to the EOS. In that context,
it has been pointed out \cite{O+98} that, unlike proton densities,          
 neutron densities are very sensitive to the EOS model. At the same time,
 precise data on neutron radii and neutron skins are not readily available
 to discriminate clearly among models.                      
Even for a nucleus such as $^{208}$Pb, for which a fairly large database
exists, determinations of the neutron skins differ considerably from 
model to model \cite{KABD}. We are presently in the process of using our EOS 
in calculations of neutron radii and neutron skins. 

The relative simplicity of a homogeneous infinite system makes
 nuclear matter calculations a convenient starting point for the 
determination of an effective interaction suitable for finite nuclei.
Together with the local density approximation, this approach has been used
extensively and with success for proton-nucleus scattering. With that in mind, we 
have taken a closer look at the scattering matrix in asymmetric matter.

Ultimately, the goal is to study 
nuclei with high levels of asymmetry, about which very little is known.
Coherent effort from both the experimental and the theoretical side is necessary 
in order to combine reliable models of the density dependence of the effective nuclear force
with reliable structure information. 
Hopefully, the RIA facility will be available in the near future to answer open
questions and help us understand the physics of the weakly bound systems that
are expected to exist at the limits of the nuclear chart.

\acknowledgements

Financial support from the U.S. \
Department of Energy under grant No.\ DE-FG03-00ER41148 is acknowledged. 
We are grateful to P. Danielewicz and W.G. Lynch for providing
the pressure contours. 

\appendix 

\section{Elimination of the angular dependence}

\subsection{Angle-averaged Pauli operator}

The definition of the Pauli operator is, for the case of two
different Fermi momenta

\begin{equation}\label{eq:fo1}
Q= \left\{ \begin{array}{lll}
0 & if & \left\{ \begin{array}{c} 
 k_{\mu} \le k^{n}_F\\
 or\\
 k_{\nu} \le k^{p}_F
\end{array} \right\}\\
\ \\
1 & if & \left\{ \begin{array}{c}
 k_{\mu}>k^{n}_F\\
 and\\
 k_{\nu}>k^{p}_F
\end{array} \right\}\\
\end{array}
\right.
\qquad\qquad (\mu \neq \nu=1,2)
\end{equation}

\ 

In the angle-average approximation, one replaces the exact $Q$ 
operator with its average over all angles for fixed $P$ and $K$.
That is, one defines 
\

\begin{equation}\label{eq:fo2}
\widetilde{Q}=<Q>=\frac{\int Q(K,P,\theta)d \Omega}{\int d \Omega}=
\frac{1}{2} \int_{\theta_{1}}^{\theta_{2}} Q(P,K,\theta) d \theta
\end{equation}

\

The variables $\vec{P},\vec{K}, \vec{k}_{\mu}$ are as 
 defined previously,
$\theta$ is the angle 
between ${\vec P}$ and ${\vec K}$, and 
$Q \neq 0$ for $\theta_{1} < \theta < \theta_{2}$.

Several cases can occur, depending 
on the values of $K\ ,P\ ,k^{p}_{F}\, and \ k^{n}_{F}$.
Those are:                            

\vspace{0.5cm}
\begin{equation}\label{eq:fo3}
\widetilde{Q}\ =\  
\left\{ \begin{array}{lc}
0 & K^{2}<\frac{1}{2}((k^{n}_{F})^{2}+(k^{p}_{F})^{2})-P^{2}\\
1 & (P-K)^{2}>(k^{n}_{F})^{2}\\
\frac{1}{4PK}
\left( (K+P)^{2}-(k^{n}_{F})^{2}
\right) & (k^{p}_{F})^{2}<(P-K)^{2}<(k^{n}_{F})^{2}\\
\frac{1}{2PK}
\left( K^{2}+P^{2}-\frac{1}{2}((k^{n}_{F})^{2}+(k^{p}_{F})^{2})
\right) & otherwise
\end{array} \right\}
\end{equation}
\vspace{0.5cm}

\subsection{Averaged center-of-mass momentum}
To simplify the integration leading to the single-particle spectrum 
and the energy per particle (see Appendices {\bf B} and {\bf C}), we introduce the average center-of-mass momentum. This is defined as 
the root-mean-square value of $P$ for two particles with 
the constraint that their relative momentum is $K$ (case $a$),    
and, in addition, that one of the particles has momentum $k_{\mu}$ (case $b$). 
\subsubsection{$K$ fixed}

Definition of average c.m. momentum (K fixed):

\begin{equation} \label{eq:ava}
P^{2}_{AV}\ =\ 
\frac
{\int^{k^{n}_{F}}_{0} d\vec k_{1}\ 
\int^{k^{p}_{F}}_{0} d\vec k_{2}
\ P^{2}\ \delta (K\ -\frac{1}{2}\ |\vec k_{1} - \vec k_{2}|)}
{\int^{k^{n}_{F}}_{0} d\vec k_{1}\ 
\int^{k^{p}_{F}}_{0} d\vec k_{2}\ 
\delta (K\ -\frac{1}{2}\ |\vec k_{1} - \vec k_{2}|)}
\end{equation}

To simplify the final expressions, we introduce the           
following notation

\begin{equation}\label{eq:not}
x=k^{n}_{F}+K\qquad y=k^{p}_{F}-K\qquad  
s=k^{n}_{F}-K\qquad  t=k^{p}_{F}+K  
\end{equation}

The final expression is then

\vspace{0.5cm}
\begin{equation}\label{eq:ava2}
P^{2}_{AV}=
\left\{
\begin{array}{lc}
\frac{3}{5}(k^{p}_{F})^{2}+K^{2} & 
\qquad\qquad\qquad\qquad\qquad\qquad\qquad\qquad\qquad K<k^{p}_{F}\\
\setlength{\unitlength}{1mm}
\begin{picture}(0,15)
\put(0,9){$\frac{8}{5}K(s^{5}+y^{5})+\frac{1}{12}(ty+sx)^{3}+
\frac{2}{3}(s^{6}+y^{6})-(ty^{5}+sx^{5})$} 
\put(0,6){\line(1,0){95}}
\put(105,6){otherwise}
\put(0,1){$\frac{8}{3}K(s^{3}+y^{3})+\frac{1}{2}(ty+sx)^{2}+
(s^{4}+y^{4})-2(ty^{3}+sx^{3})$} 
\end{picture} & \ \\
0 & 
\qquad\qquad\qquad\qquad\qquad\qquad
\qquad\qquad\qquad (k^{n}_{F}+k^{p}_{F})<2K
\end{array}
\right.
\end{equation}


\subsubsection{$K$ and $k_{\mu}$ fixed}

Definition of average c.m. momentum (K and $k_{\mu}$ fixed):

\begin{equation} \label{eq:ava1}
{\widetilde P}^{2}_{AV}\ =\ 
\frac
{\int_{\Omega_{\mu}} d\Omega_{\mu}\ 
\int^{k^{i}_{F}}_{0} d\vec k_{\nu}
\ P^{2}\ \delta (K\ -\frac{1}{2}\ |\vec k_{\mu} - \vec k_{\nu}|)}
{\int_{\Omega_{\mu}} d\Omega_{\mu}\ 
\int^{k^{i}_{F}}_{0} d\vec k_{2\nu}\ 
\delta (K\ -\frac{1}{2}\ |\vec k_{\mu} - \vec k_{\nu}|)}
\end{equation}

\begin{tabular}{|p{2cm}|}
\hline
$k_{\mu}\ <\ k^{i}_{F}$
\\
\hline
\end{tabular}

\vspace{0.5cm}

\begin{equation}\label{eq:ava3}
\widetilde{P}^{2}_{AV}\ =\  
\left\{ \begin{array}{lc}
(k_{\mu})^{2}+K^{2}
& \qquad 2K<k^{i}_{F}-k_{\mu}\\
\frac{1}{4} \left( 3(k_{\mu})^{2}+(k^{i}_{F})^{2}-4 K k_{\mu}
\right) 
& \qquad k^{i}_{F}-k_{\mu}<2K<k^{i}_{F}+k_{\mu}\\
0 & \qquad otherwise
\end{array} \right\}
\end{equation}

\begin{tabular}{|p{2cm}|}
\hline
$k_{\mu}\ >\ k^{i}_{F}$
\\
\hline
\end{tabular}

\vspace{0.5cm}

\begin{equation}\label{eq:ava4}
\widetilde{P}^{2}_{AV}\ =\  
\left\{ \begin{array}{lc}
\frac{1}{4} \left( 3(k_{\mu})^{2}+(k^{i}_{F})^{2}-4 K k_{\mu}
\right) 
& \qquad k_{\mu}-k^{i}_{F}<2K<k^{i}_{F}+k_{\mu}\\
0 & \qquad otherwise
\end{array} \right\}
\end{equation}

where $\mu=1,2$ and i$=$n or p.

\section{Nuclear matter single-particle potential}

The single-particle potential in nuclear matter,             
 $U_{i}$ (i$=$n or p), is defined in the usual way \cite{BBP}

\begin{displaymath}
U_{i}(k_{\mu})=\sum_{j<A}<\mu_{i} \nu_{j}|
G^{ij}|\mu_{i} \nu_{j} -\nu_{j} \mu_{i}>=
\end{displaymath}

\begin{equation} \label{eq:nmp1}
\sum_{j_{1}<N}<\mu_{i} \nu_{j_{1}}|
G^{in}|\mu_{i} \nu_{j_{1}} - \nu_{j_{1}} \mu_{i}>+
\sum_{j_{2}<Z}<\mu_{i} \nu_{j_{2}}|
G^{ip}|\mu_{i} \nu_{j_{2}}- \nu_{j_{2}} \mu_{i}>
\end{equation}

The explicit evaluation of this expression in 
partial-wave decomposition yields 

\begin{displaymath}
U_{i}(k_{\mu})= \frac{8\lambda}{ \pi k_{\mu}} \sum_{L, \alpha} (2J+1)
{\mathcal{T}}^{T}_{in} \Bigg[ \int_{0}^{\frac{k^{n}_{F}-k_{\mu}}{2}}
K_{0} dK_{0} 
\int_{|k_{\mu}-K_{0}|}^{k_{\mu}+K_{0}}P dP G^{\alpha,in}_{LL}
(P;K_{0},K_{0})
\end{displaymath}

\begin{equation} \label{eq:nmp8} 
+ \int_{\frac{k^{n}_{F}-k_{\mu}}{2}}^{\frac{k^{n}_{F}+k_{\mu}}{2}}
K_{0} dK_{0} 
\int_{|k_{\mu}-K_{0}|}
^{\sqrt{\frac{k^{2}_{\mu}+(k^{n}_{F})^{2}}{2}-K^{2}_{0}}}P dP 
G^{\alpha,in}_{LL}(P;K_{0},K_{0})\Bigg]
+
\left( \begin{array}{c}
n\ \longleftrightarrow \ p\\
k^{n}_{F}\ \longleftrightarrow \ k^{p}_{F}
\end{array} \right)
\end{equation} 

($\alpha$ stands for the quantum numbers J, S and T).

Using the averaged-momentum 
approximation (see Appendix {\bf A}),  

\begin{equation} \label{eq:comapp2}
G^{\alpha,ij}_{LL}(P;K_{0},K_{0})
\approx
G^{\alpha,ij}_{LL}({\tilde P_{AV}};K_{0},K_{0}), 
\end{equation}

allows to simplify the integral over P, 
with the final result being

\begin{displaymath}
U_{i}(k_{\mu})= \frac{8\lambda}{ \pi k_{\mu}} \sum_{L, \alpha} (2J+1)
{\mathcal{T}}^{T}_{in} \Bigg[ \int_{0}^{\frac{k^{n}_{F}-k_{\mu}}{2}}
K_{0} dK_{0} 
G^{\alpha,in}_{LL}({\tilde P_{AV}};K_{0},K_{0})
\int_{|k_{\mu}-K_{0}|}^{k_{\mu}+K_{0}}P dP 
\end{displaymath}

\begin{displaymath}
+ \int_{\frac{k^{n}_{F}-k_{\mu}}{2}}^{\frac{k^{n}_{F}+k_{\mu}}{2}}
K_{0} dK_{0} 
G^{\alpha,in}_{LL}({\tilde P_{AV}};K_{0},K_{0})
\int_{|k_{\mu}-K_{0}|}
^{\sqrt{\frac{k^{2}_{\mu}+(k^{n}_{F})^{2}}{2}-K^{2}_{0}}}P dP 
\Bigg]
+
\left( \begin{array}{c}
n\ \longleftrightarrow \ p\\
k^{n}_{F}\ \longleftrightarrow \ k^{p}_{F}
\end{array} \right)
\end{displaymath}

\begin{displaymath}
= \frac{16\lambda}{ \pi } \sum_{L, \alpha} (2J+1)
{\mathcal{T}}^{T}_{in} \Bigg[ \int_{0}^{\frac{k^{n}_{F}-k_{\mu}}{2}}
K^{2}_{0} dK_{0} 
G^{\alpha,in}_{LL}({\tilde P_{AV}};K_{0},K_{0})
\end{displaymath}

\begin{equation} \label{eq:nmp9} 
+ \frac{1}{2k_{\mu}}
\int_{\frac{k^{n}_{F}-k_{\mu}}{2}}^{\frac{k^{n}_{F}+k_{\mu}}{2}}
K_{0} dK_{0} 
\Big[ \frac{(k^{n}_{F})^{2}-k_{\mu}^{2}}{4}-K_{0}(K_{0}-k_{\mu})
\Big]
G^{\alpha,in}_{LL}({\tilde P_{AV}};K_{0},K_{0})
\Bigg]
+
\left( \begin{array}{c}
n\ \longleftrightarrow \ p\\
k^{n}_{F}\ \longleftrightarrow \ k^{p}_{F}
\end{array} \right)
\end{equation} 

The coefficients 
${\mathcal{T}}^{T}_{ij}$ contain the isospin dependence and are
equal to 

\begin{center}
${\mathcal{T}}^{T=0,1}_{np}=$ 
${\mathcal{T}}^{T=0,1}_{pn}=\frac{1}{2}$ ;\qquad
${\mathcal{T}}^{0}_{nn}=$ 
${\mathcal{T}}^{0}_{pp}=0$ ;\qquad

${\mathcal{T}}^{1}_{nn}=$ 
${\mathcal{T}}^{1}_{pp}=1$ 
\end{center}

\section{Energy per nucleon in nuclear matter}

The energy per neutron/proton is:       

\begin{equation}\label{eq:spn}
\bar{e}_{n}=
<T_{n}> + \frac{1}{2N} \sum_{j<A} \sum_{i<N}
<\mu_{i}\nu_{j}|G_{ij}|\mu_{i}\nu_{j}-\nu_{j}\mu_{i}>
\end{equation}

\begin{equation}\label{eq:spp}
\bar{e}_{p}=
<T_{p}> + \frac{1}{2Z} \sum_{j<A} \sum_{i<Z}
<\mu_{i}\nu_{j}|G_{ij}|\mu_{i}\nu_{j}-\nu_{j}\mu_{i}>
\end{equation}

which, in view of                                                       
Eq.(\ref{eq:nmp1}), can be written as 

\begin{equation}\label{eq:sin1}
\bar{e}_{n}=
<T_{n}> + \frac{1}{2N} \sum_{i<N} U_{i}(k_{\mu})
\end{equation}

\begin{equation}\label{eq:sin2}
\bar{e}_{p}=
<T_{p}> + \frac{1}{2Z} \sum_{i<Z} U_{i}(k_{\mu})
\end{equation}

By applying again the partial-wave decomposition and
the average momentum approximation

\begin{equation} \label{eq:comapp3}
G^{\alpha,ij}_{LL}(P;K_{0},K_{0})
\approx
G^{\alpha,ij}_{LL}(P_{AV};K_{0},K_{0})
\end{equation}

the energy per neutron/proton is finally 

\begin{displaymath}
\bar{e}_{i}=
<T_{i}> + 
\frac{6 \lambda}{\pi (k^{i}_{F})^{3}}\sum_{L,\alpha} (2J+1)
\left[
\frac{4}{3}(k^{p}_{F})^{3} {\mathcal{T}}^{T}_{pn}
\int^{\frac{k^{n}_{F}-k^{p}_{F}}{2}}
_{0}
K^{2}_{0}dK_{0}
G^{\alpha(np)}_{LL}(P_{AV};K_{0},K_{0})+
\right.
\end{displaymath}

\begin{displaymath}
2 {\mathcal{T}}^{T}_{pn}\int^{\frac{k^{n}_{F}+k^{p}_{F}}{2}}
_{\frac{k^{n}_{F}-k^{p}_{F}}{2}}
K_{0}^{2}dK_{0}
\left[
\frac{(k^{n}_{F})^{3}+(k^{p}_{F})^{3}}{3}-
\frac{((k^{n}_{F})^{2}-(k^{p}_{F})^{2})^{2}}{16K_{0}}+
\frac{K^{3}_{0}}{3}-
\frac{K_{0}}{2}((k^{n}_{F})^{2}+(k^{p}_{F})^{2})
\right]
\end{displaymath}

\begin{equation}\label{eq:be6}
G^{\alpha(np)}_{LL}(P_{AV};K_{0},K_{0})+
\left.
\frac{2}{3}(k^{i}_{F})^{3}
 {\mathcal{T}}^{T}_{ii}
\int^{k^{i}_{F}}
_{0}
K_{0}^{2}dK_{0}
\left[
2-3 \frac{K_{0}}{k^{i}_{F}}+
(\frac{K_{0}}{k^{i}_{F}})^{3}
\right]
G^{\alpha(ii)}_{LL}(P_{AV};K_{0},K_{0})
\right]
\end{equation}

where, again, i $=$ n or p.

\newpage 

\begin{figure}
\begin{center}
\vspace*{1cm}
\hspace*{-1cm}
\psfig{figure=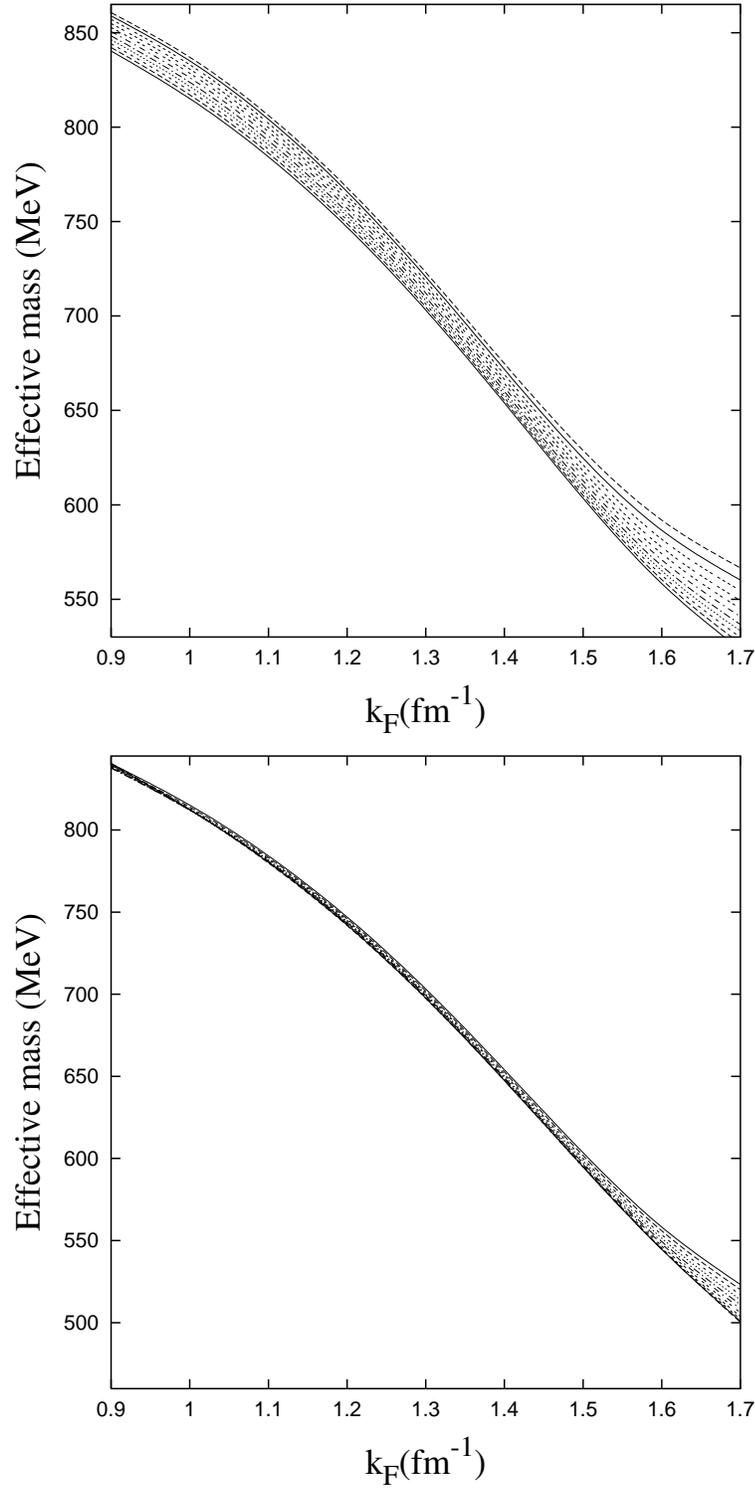,height=10cm}
\vspace*{9.0cm}
\caption{Upper panel: neutron effective mass as a function of the Fermi
momentum and for increasing values of $\alpha$ between 0 and 1.
The upper curve corresponds to $\alpha = 1$. Lower panel: proton effective mass. 
The upper curve corresponds to $\alpha = 0$.}
\label{one}
\end{center}
\end{figure}

\newpage
\begin{figure}
\begin{center}
\vspace*{1cm}
\hspace*{-1cm}
\psfig{figure=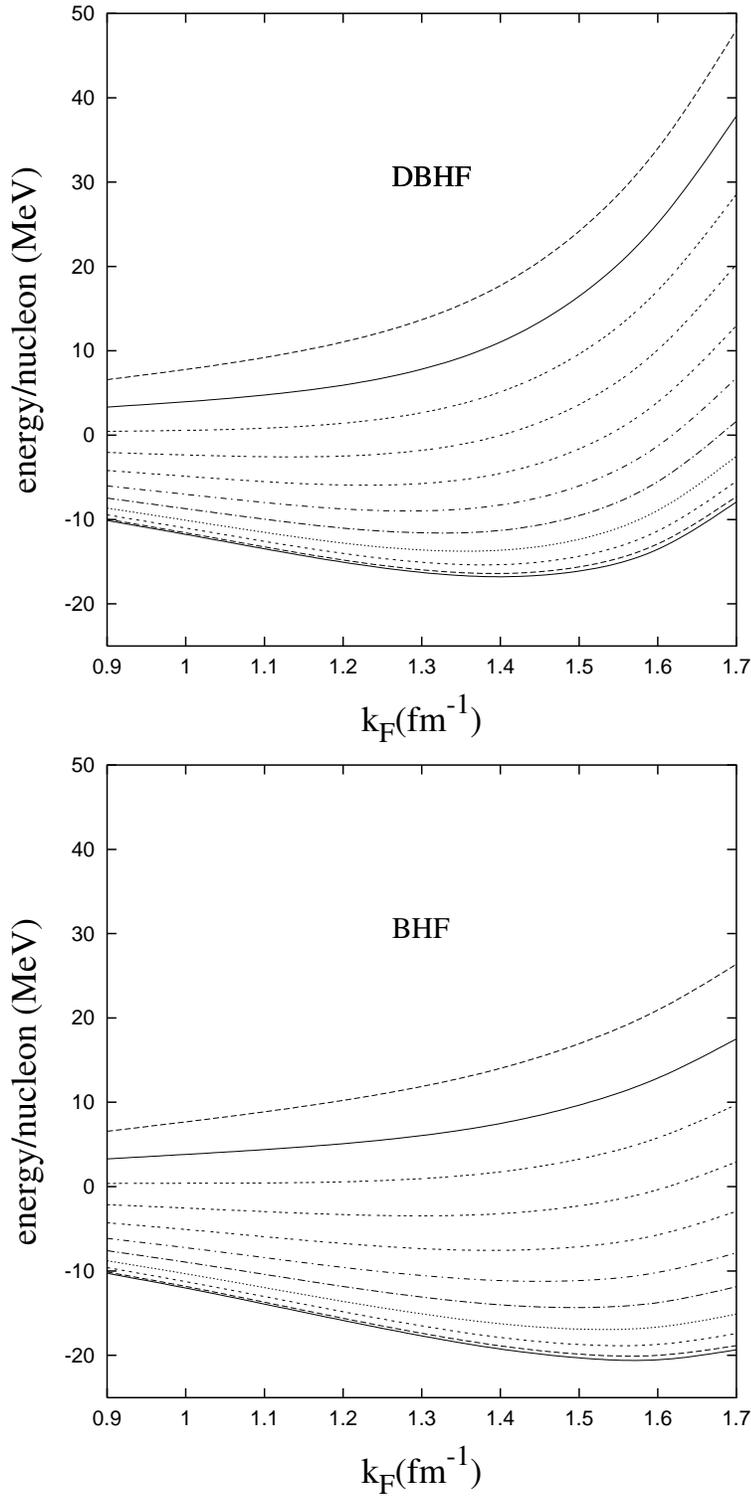,height=10cm}
\vspace*{9.0cm}
\caption{Upper panel: energy per nucleon as a function of the Fermi momentum at
differente values of the asymmetry parameter (in steps of 0.1)
from symmetric matter (lowest curve) to neutron matter (highest curve).
The predictions are obtained from DBHF calculations. Lower panel: corresponding predictions
from BHF calculations. 
}
\label{two}
\end{center}
\end{figure}

\newpage
\begin{figure}
\begin{center}
\vspace*{1cm}
\hspace*{-1cm}
\psfig{figure=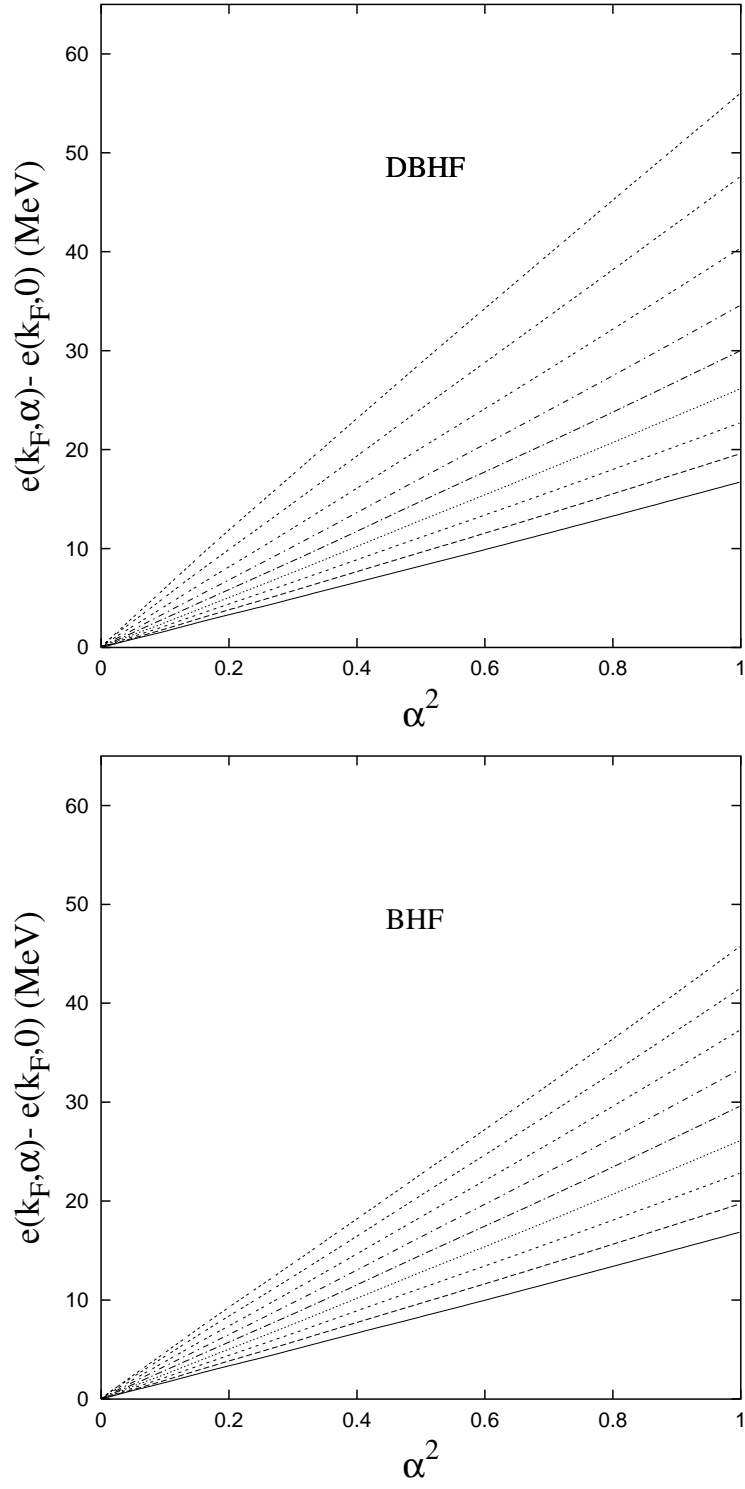,height=10cm}
\vspace*{9.0cm}
\caption{Upper panel: the left-hand side of Eq.~(20) versus $\alpha ^2$ (DBHF
model) for increasing values of the Fermi momentum from 0.9 $fm^{-1}$ (lowest curve)
to 1.7 $fm^{-1}$. Lower panel: corresponding predictions from BHF calculations.} 
\label{three}
\end{center}
\end{figure}

\newpage
\begin{figure}
\begin{center}
\vspace*{1cm}
\hspace*{-1cm}
\psfig{figure=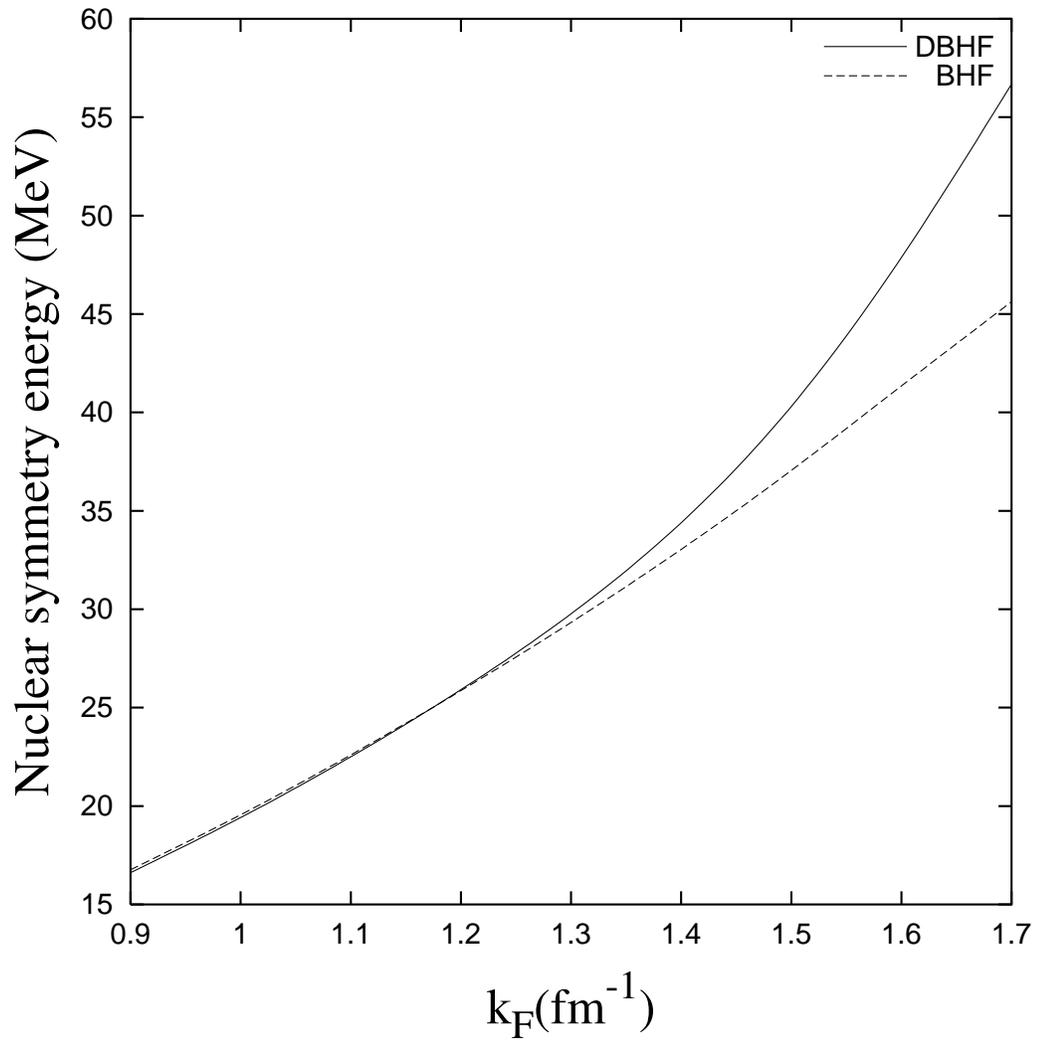,height=14cm}
\vspace*{2.5cm}
\caption{Nuclear symmetry energy as a function of the Fermi momentum.
The solid line is the prediction from DBHF calculations, while the
dashed line is obtained with the conventional Brueckner approach.}
\label{four}
\end{center}
\end{figure}

\newpage
\begin{figure}
\begin{center}
\vspace*{1cm}
\hspace*{-1cm}
\psfig{figure=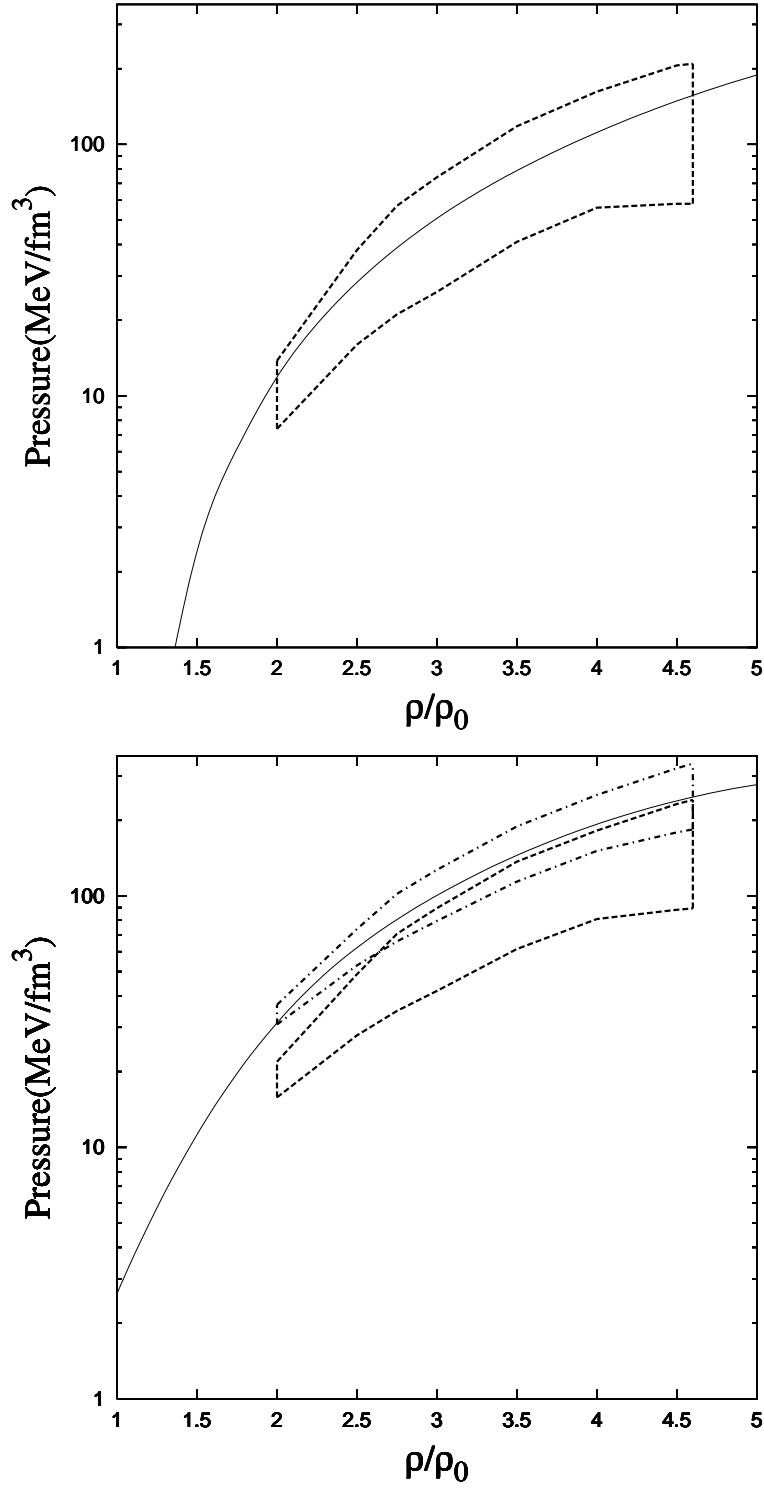,height=10cm}
\vspace*{9.0cm}
\caption{Upper panel: pressure (in MeV/$fm^3$) in symmetric matter as a function of 
 density (in units of saturation density). The area within the 
dashed lines indicates pressure values consistent with the data. Lower panel: 
pressure in neutron matter. The two contours correspond to two parametrizations of the 
asymmetry term used in the analysis, as explained in the text.                  
The predictions are from DBHF calculations.}                   
\label{five}
\end{center}
\end{figure}

\newpage
\begin{figure}
\begin{center}
\vspace*{1cm}
\hspace*{-1cm}
\psfig{figure=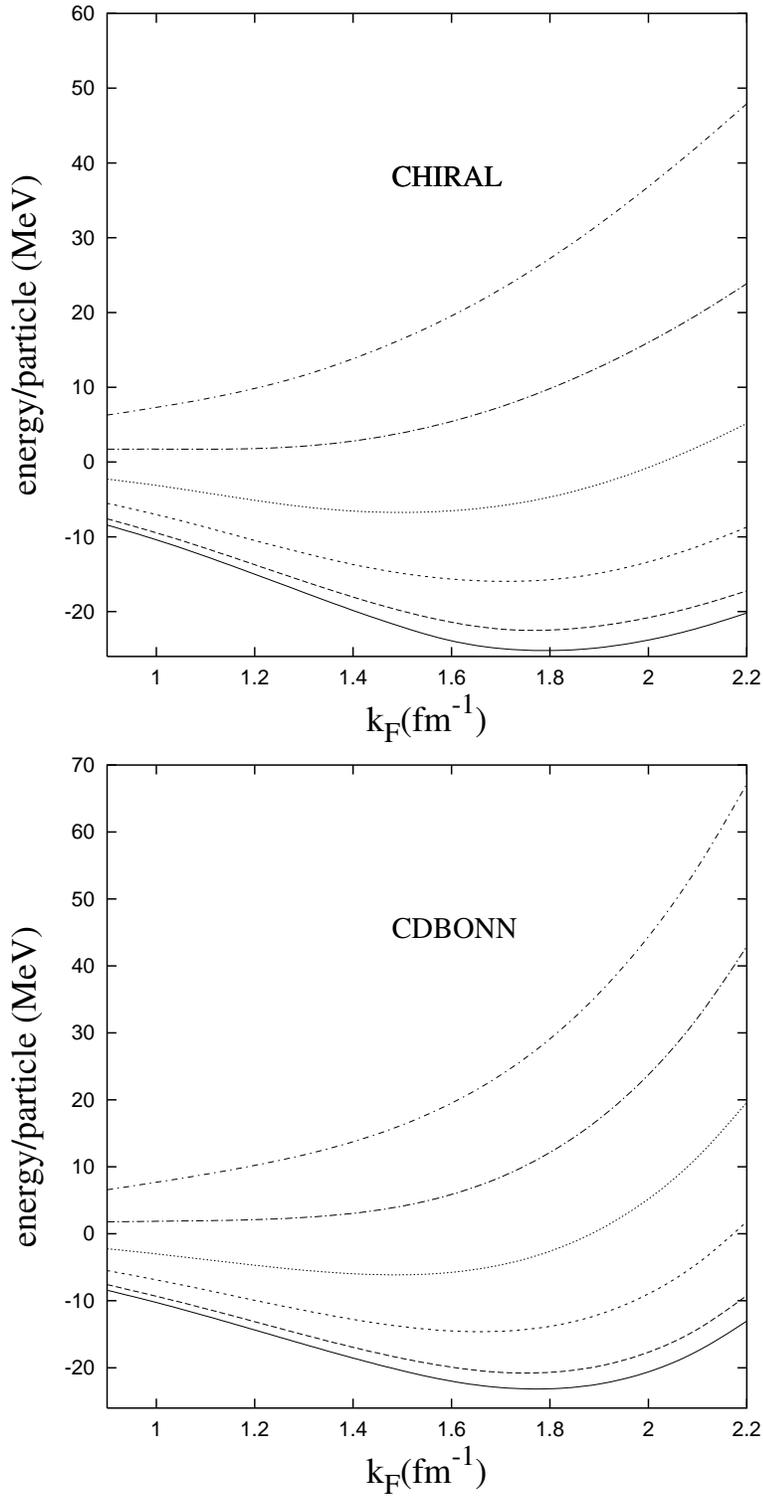,height=10cm}
\vspace*{9.0cm}
\caption{Upper panel: energy per nucleon as a function of the Fermi momentum at
differente values of the asymmetry parameter (in steps of 0.2)
from symmetric matter (lowest curve) to neutron matter (highest curve).
The predictions are obtained from BHF calculations with the chiral potential. Lower panel: corresponding predictions
with CD-Bonn.                  
}
\label{six}
\end{center}
\end{figure}

\newpage
\begin{figure}
\begin{center}
\vspace*{1cm}
\hspace*{-1cm}
\psfig{figure=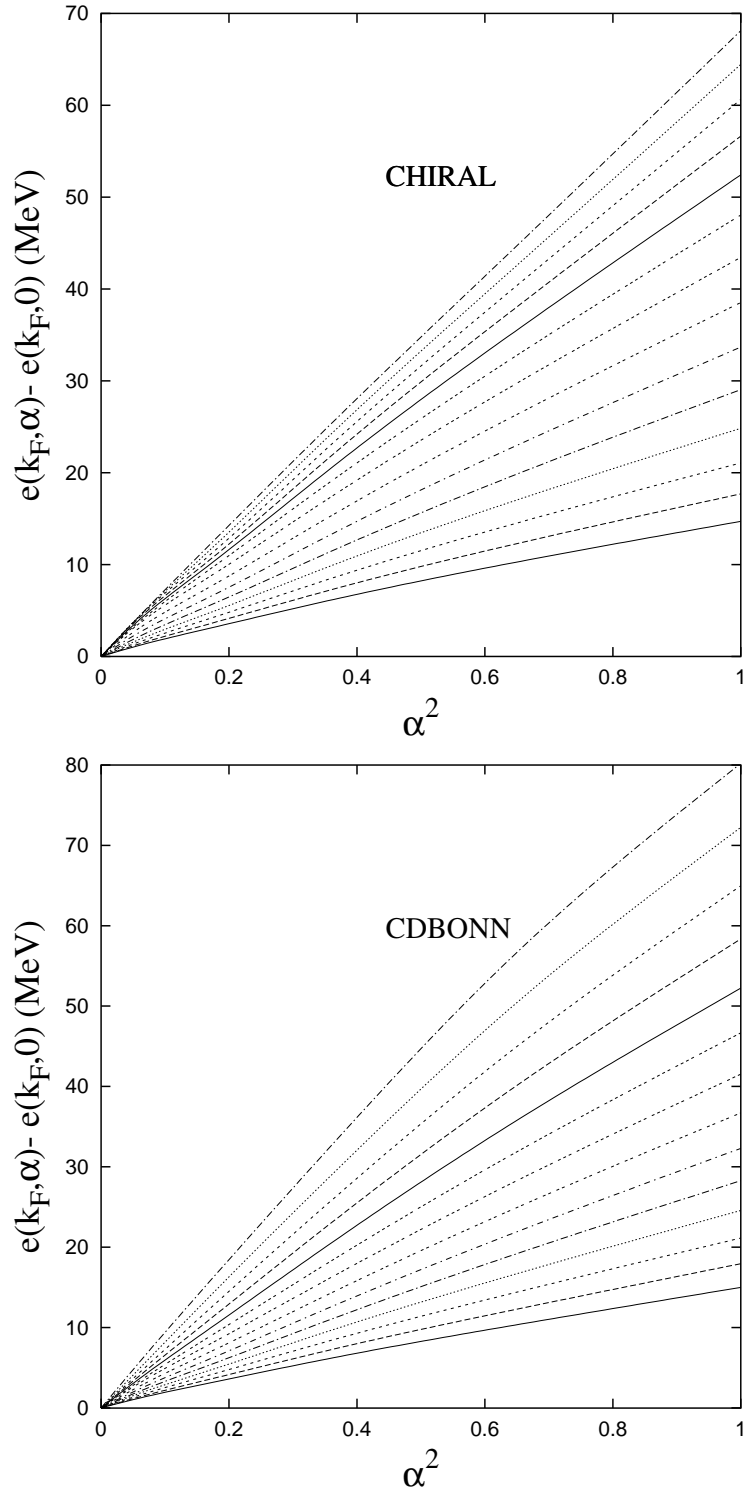,height=10cm}
\vspace*{9.0cm}
\caption{Upper panel: the left-hand side of Eq.~(20) versus $\alpha ^2$ (obtained with the chiral potential)
 for increasing values of the Fermi momentum from 0.9 $fm^{-1}$ (lowest curve)
to 2.2 $fm^{-1}$. Lower panel: corresponding predictions with CD-Bonn.} 
\label{seven}
\end{center}
\end{figure}

\newpage
\begin{figure}
\begin{center}
\vspace*{1cm}
\hspace*{-1cm}
\psfig{figure=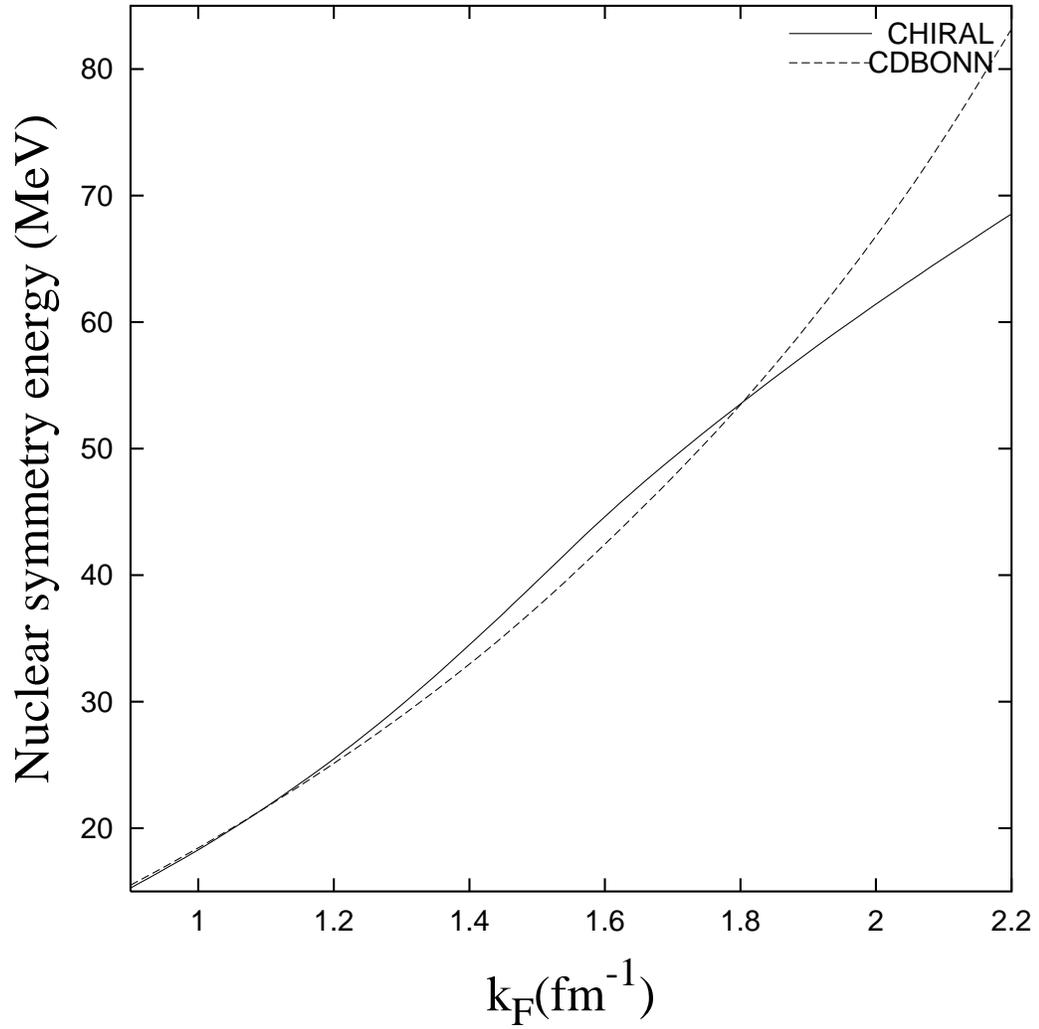,height=14cm}
\vspace*{2.5cm}
\caption{Nuclear symmetry energy as a function of the Fermi momentum.
The solid line is the prediction obtained with the chiral potential, while the
dashed line is obtained with CD-Bonn.}
\label{eight}
\end{center}
\end{figure}

\newpage
\begin{figure}
\begin{center}
\vspace*{1cm}
\hspace*{-1cm}
\psfig{figure=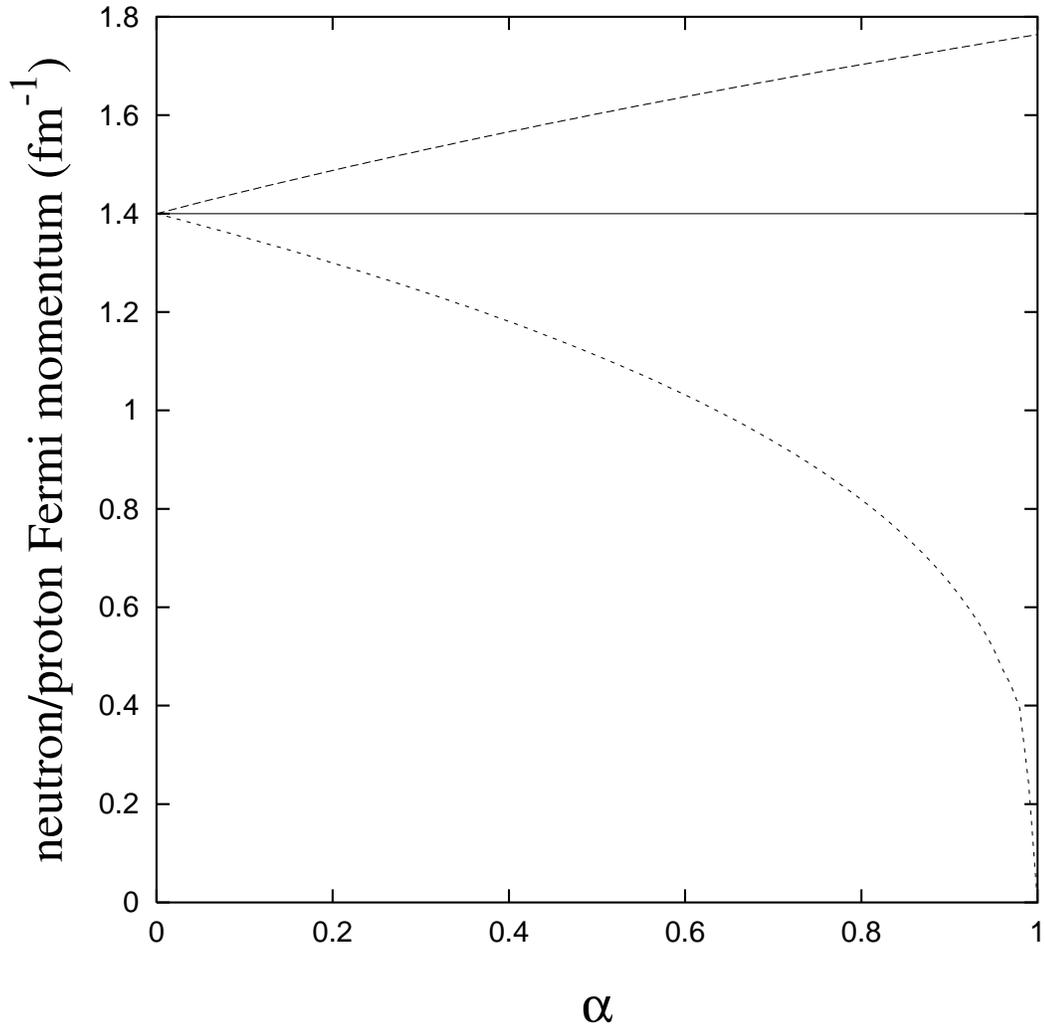,height=14cm}
\vspace*{2.5cm}
\caption{The proton (lower line) and neutron (upper line) Fermi momenta 
as a function of the asymmetry parameter for fixed value of the (average) Fermi momentum (solid line). 
} 
\label{nine}
\end{center}
\end{figure}

\newpage
\begin{figure}
\begin{center}
\vspace*{1cm}
\hspace*{-1cm}
\psfig{figure=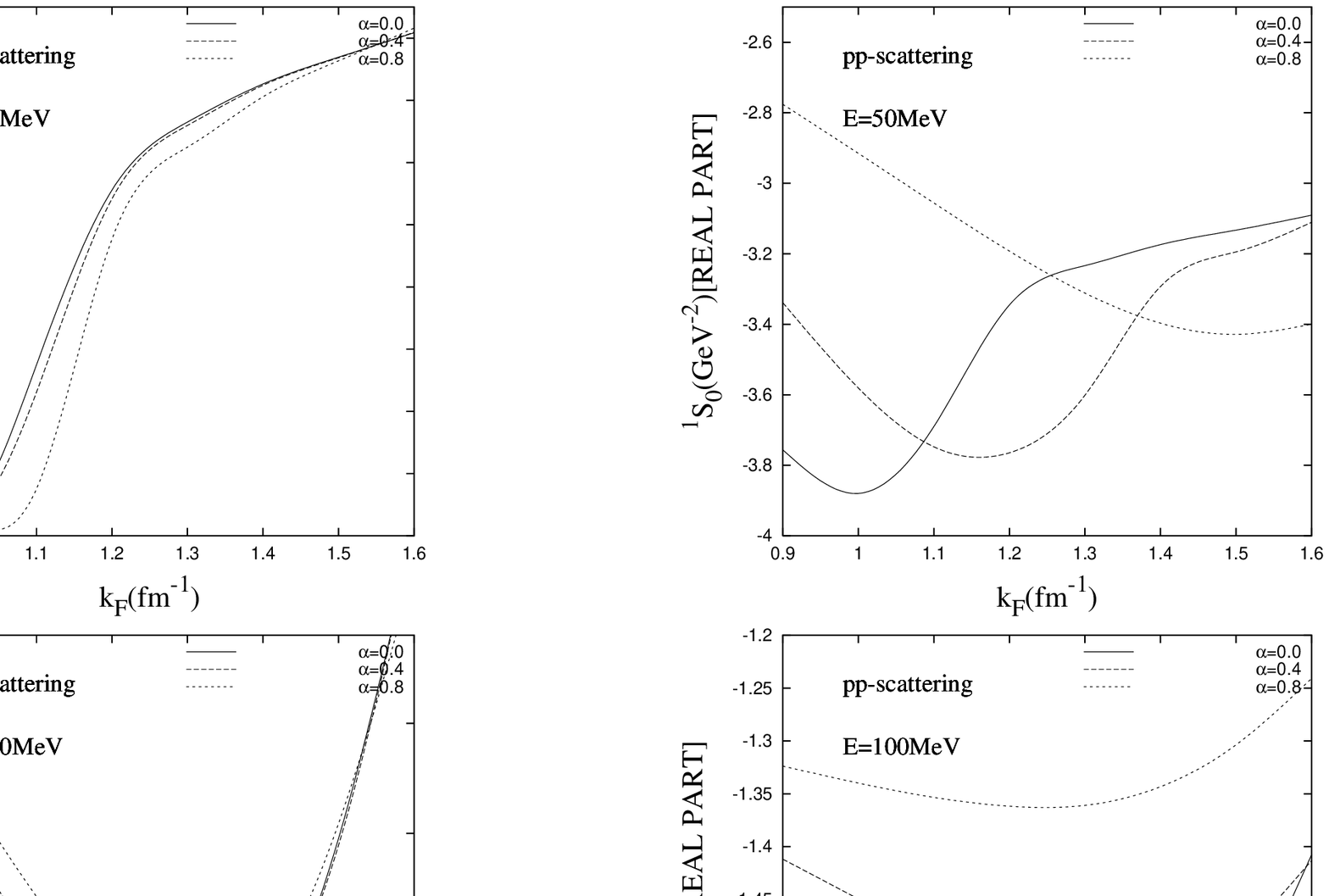,height=7cm}
\vspace*{13.0cm}
\caption{ Real part of the $np$ and $pp$ $^{1}S_{0}$ matrix element as a function 
of the Fermi momentum at different energies and for different levels of asymmetry.} 
\label{ten}
\end{center}
\end{figure}

\newpage
\begin{figure}
\begin{center}
\vspace*{1cm}
\hspace*{-1cm}
\psfig{figure=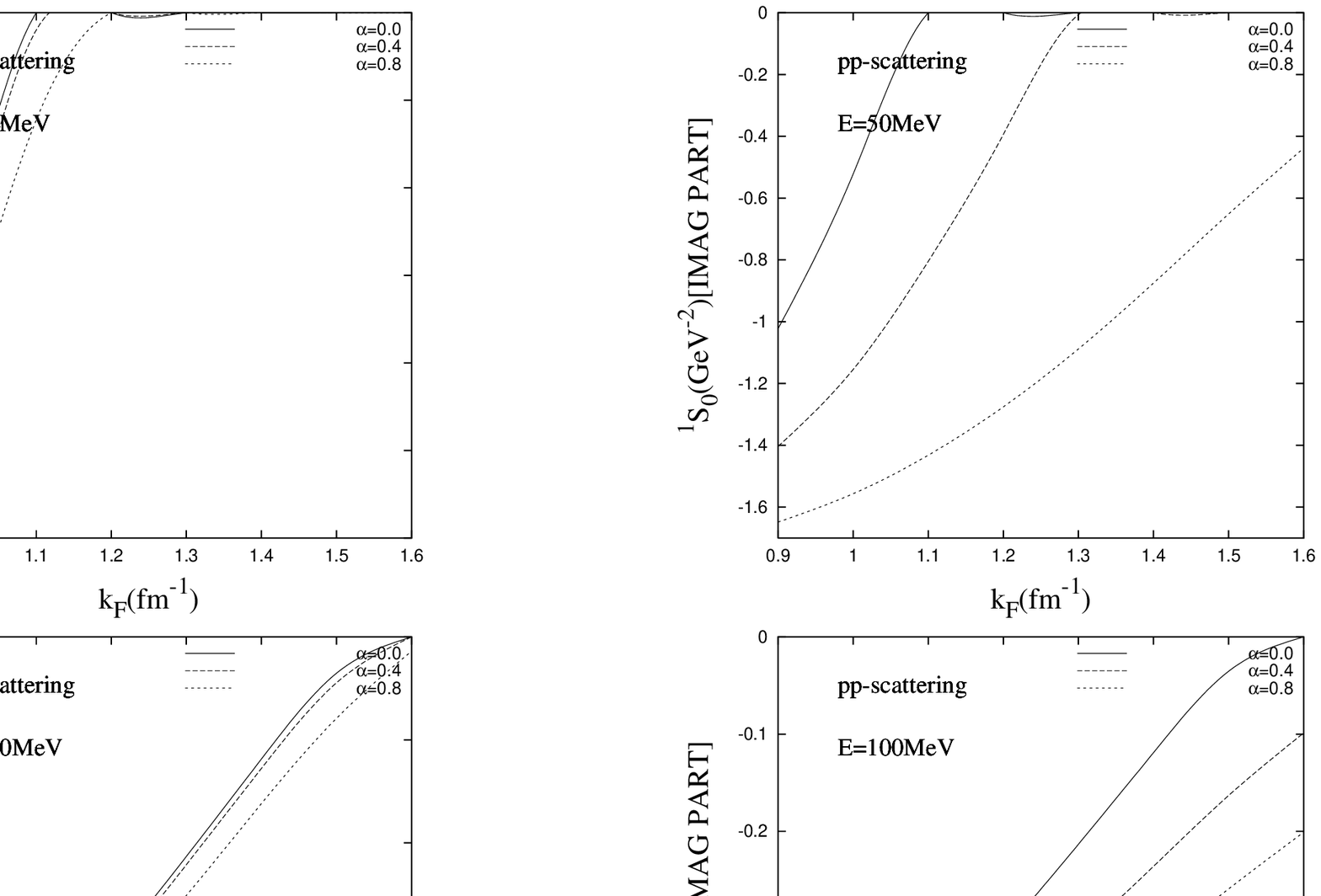,height=7cm}
\vspace*{13.0cm}
\caption{ Imaginary part of the $np$ and $pp$ $^{1}S_{0}$ matrix element as a function 
of the Fermi momentum at different energies and for different levels of asymmetry.} 
\label{eleven}
\end{center}
\end{figure}

\newpage
\begin{figure}
\begin{center}
\vspace*{1cm}
\hspace*{-1cm}
\psfig{figure=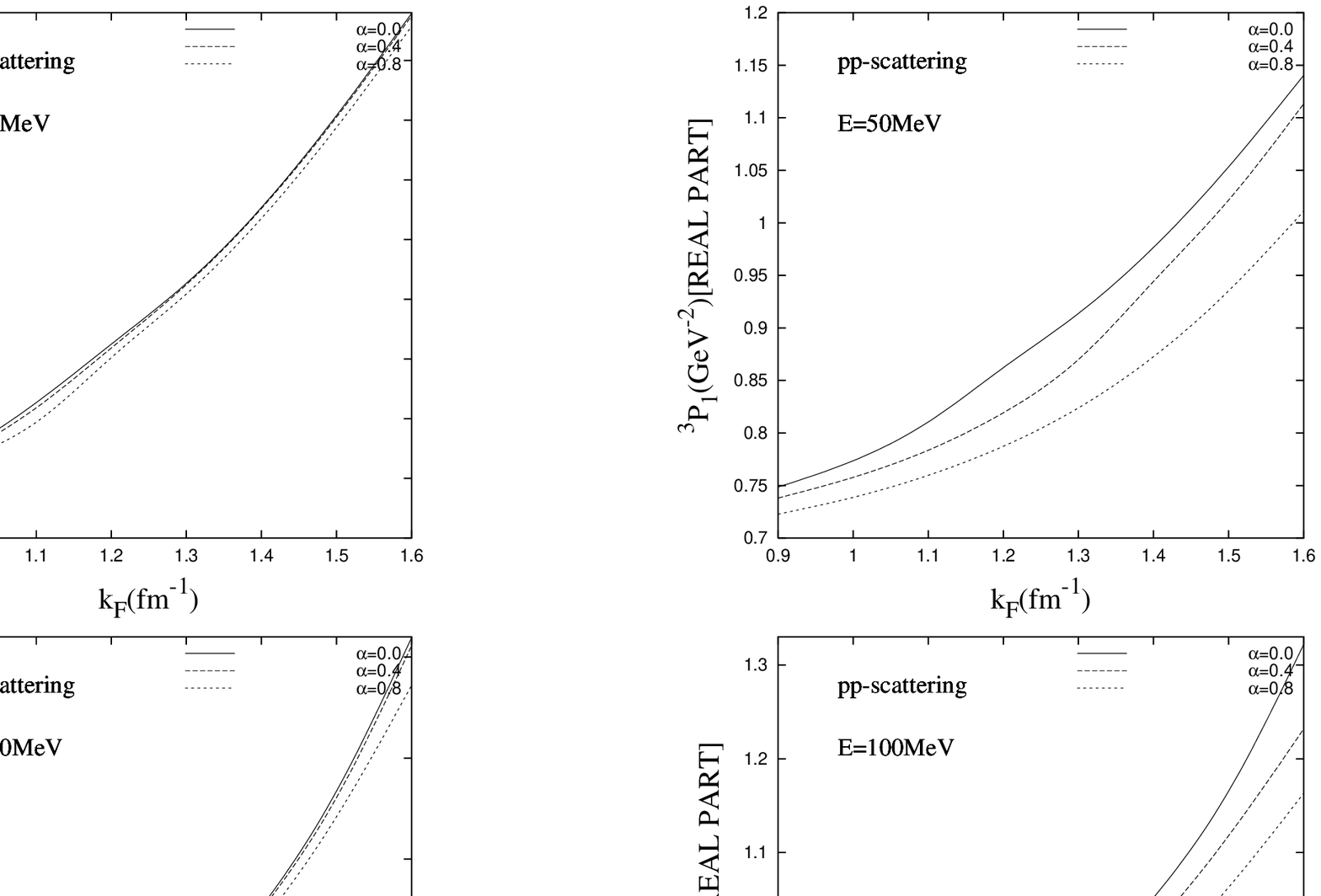,height=7cm}
\vspace*{13.0cm}
\caption{ As in Fig.~10, for the  $^{3}P_{1}$ matrix element.}             
\label{twelve}
\end{center}
\end{figure}

\newpage
\begin{figure}
\begin{center}
\vspace*{1cm}
\hspace*{-1cm}
\psfig{figure=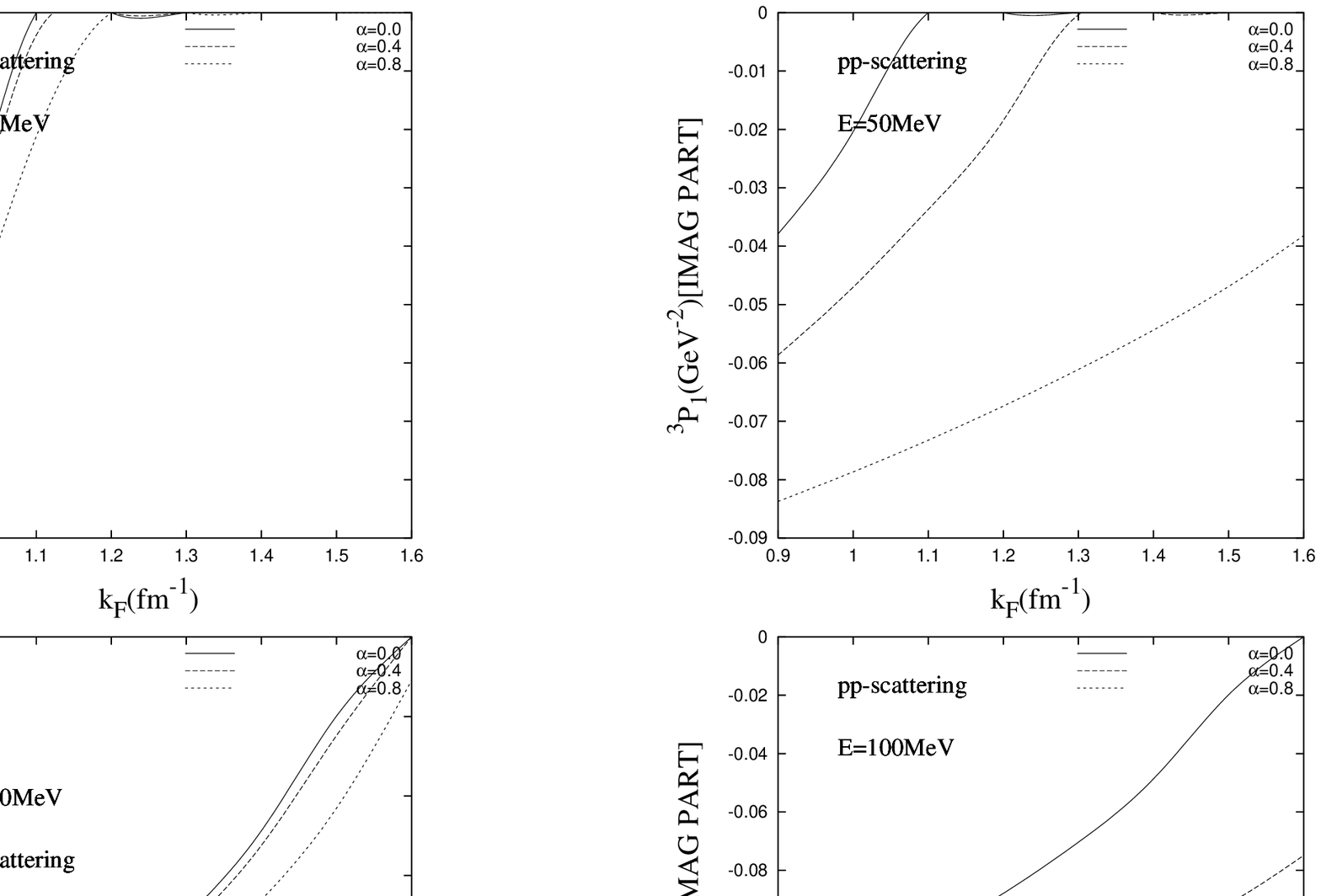,height=7cm}
\vspace*{13.0cm}
\caption{ As in Fig.~11, for the  $^{3}P_{1}$ matrix element.}             
\label{thirteen}
\end{center}
\end{figure}

\end{document}